\newcommand{\bee}{\begin{equation}}
\newcommand{\ene}{\end{equation}}
\newcommand{\beea}{\begin{eqnarray}}
\newcommand{\enea}{\end{eqnarray}}
\newcommand{\lt}{\left}
\newcommand{\rt}{\right}
\documentclass[pre,twocolumn,showpacs,amsmath,amssymb]{revtex4}
\usepackage[]{graphicx}
\usepackage{color}

\begin{document}
\draft

%\begin{linenumbers}
%\color{red}

\title{Dust interferometers in plasmas}

\author{M. Chaudhuri$^{1*}$, V. Nosenko$^2$, H. M. Thomas$^2$}
\affiliation{$^1$School of Engineering and Applied Sciences (SEAS), Harvard University, USA}
\affiliation{$^2$Forschungsgruppe Komplexe Plasmen,
Deutsches Zentrum f\"{u}r Luft- und Raumfahrt, D-82234, We{\ss}ling, Germany}
%\author{G. E. Morfill}
%\affiliation{Max-Planck-Institut f\"ur extraterrestrische Physik, D-85741 Garching, Germany}
%\affiliation{Centre for Plasma Science and Technology, BMSTU, Moscow, Russia}

\begin{abstract}
An interferometric imaging technique has been proposed to instantly measure the diameter of individual spherical dust particles suspended in a gas discharge plasma. The technique is based on the defocused image analysis of both spherical particles and their binary agglomerates. Above a critical diameter, the defocused images of spherical particles contain stationary interference fringe patterns and the fringe number increases with particle diameters. Below this critical diameter, the particle size has been measured using the rotational interference fringe patterns which appear only on the defocused images of binary agglomerates. In this case, a lower cut-off limit of particle diameter has been predicted, below which no such rotational fringe patterns are observed for the binary agglomerates. The method can be useful as a diagnostics for complex plasma experiments on earth as well as under microgravity condition.
\end{abstract}

\pacs{07.60.Ly, 42.25.Hz, 52.27.Lw, 52.70.Kz, 64.70.Dv, 95.55.Br}

\maketitle

%\section{Introduction}

%Soft matter is a rapidly growing interdisciplinary research covering a wide range of topics from soft nanotechnology and self-assembly (nanostructured polymeric materials, nanocomposites, molecular self-organisation, encapsulation, etc.), bulk soft matter assemblies (colloids, polymers, gels, vescicles, films, liquid crystal, etc.), to complex biological systems (biomacromolecules and biopolymers, membranes, biocomposites, biomimetic materials, etc.)~\cite{Jones,Larson,Poon}. 
%They provide a fascinating testing ground to investigate fundamental questions in condensed matter physics, as for example, many-body statistical physics, rheology, hydrodynamics, topological defects, phase transition, jamming transition and many more. Soft materials have also important industrial applications in photonics and lithography, in high-tech ceramics and in biochemical sensing. 
Understanding strongly correlated phenomena such as crystal and liquid structures, melting dynamics, crystallization, homogeneous nucleation, dendrites, glass transition, etc. in a classical many body systems are outstanding topics of practical importance in material science~\cite{Chaikin:Lubensky,kittel,Hansen:McDonald,larson}. Colloids have long been used as a model system to investigate such processes where particles of different shapes (sphere, cube, ellipsoid, agglomerates, etc.) can be synthesized based on experimental requirements~\cite{lowen1994,biben1994,Weeks:2000aa,Gasser:2001aa,Dullens2006,Kawasaki:2010aa,Peng-Tan:2014aa}. The surfaces of colloidal particles can be treated chemically to explore wide range of inter-particle interactions (repulsive to attractive, hard sphere to ultra-soft) and associated tunable, collective, self-organized processes. Different high resolution imaging techniques, such as laser scanning confocal microscope (LSCM), electron microscopy or atomic force microscopy (AFM) techniques can be used to measure size and shape of colloidal particles precisely. Other technique such as dynamic light scattering has also been used to measure hydrodynamic size of the particles in a dilute solution. In recent times, it is revealed that colloids share unique complementary features with complex (dusty) plasma which is being considered as the plasma state of soft matter~\cite{Morfill_RMP,Chaudhuri_SM,Morfill:lowen}. However, unlike colloids where the particle dynamics is over damped due to viscous solvents, the highly charged solid particles in complex plasmas levitate in the background of weakly ionized gas~\cite{Thomas_PRL,Chu_PRL,Thomas-Nature-1996}. Basic understanding of plasma-particle interactions are essential to tune inter-particle interactions and relevant self-organized collective phenomena in complex plasmas~\cite{Fortov2005PR,vladimirov-ost-phys.rep-2004,Konopka2000,lampe-gos-ste-phys.plasmas-2003,Chaudhuri2007a,Chaudhuri2008a,ChaudhuriIEEE2010}. The background neutral gas pressure can be controlled precisely to achieve almost undamped particle dynamics which makes complex plasma a unique model system to explore classical many body phenomena at the ``atomistic'' level~\cite{Morfill:lowen,Morfill:pop}. Different types of unique experiments have been performed at ground based laboratories on earth~\cite{Quinn_PhysRevE.64.051404,Zuzic:PhysRevLett.85.4064,Nosenko_PhysRevLett.93.155004,Nosenko_PhysRevLett.103.015001,Hartmann_PhysRevLett.105.115004,Feng_PhysRevLett.104.165003,Nosenko_PhysRevLett.99.025002,Knapek_PhysRevLett.98.015004,Nunomura_PhysRevLett.95.025003,Nosenko_PhysRevLett.93.155004,Liu_PhysRevLett.100.055003,Juan_PhysRevE.64.016402}, as well as under microgravity condition onboard ``International Space Station (ISS)''~\cite{Morfill:PhysRevLett.83.1598,Kretschmer:PhysRevE.71.056401,Zuzic:Naturephysics,Thomas:NJP,Schwabe:EPL}. Typically experimental data in complex plasmas is analyzed by using standard particle location and tracking methods on sequence of images obtained by video microscopy technique. Several other techniques with different features have also been used, such as, Particle-Image-Velocimetry (PIV)~\cite{Ed_Thomas_10.1063/1.873544,Ed_Thomas_10.1063/1.2174831}, digital in-line holography~\cite{Kroll_10.1063/1.2932109}, color gradient method~\cite{Annaratone_PPCF}, stereoscopy~\cite{kading:090701}, etc. However to the best of our knowledge, there exists no technique in complex plasmas to identify individual particles shape and size instantly during experiments.

\begin{figure}[h]
\includegraphics[width=\linewidth]{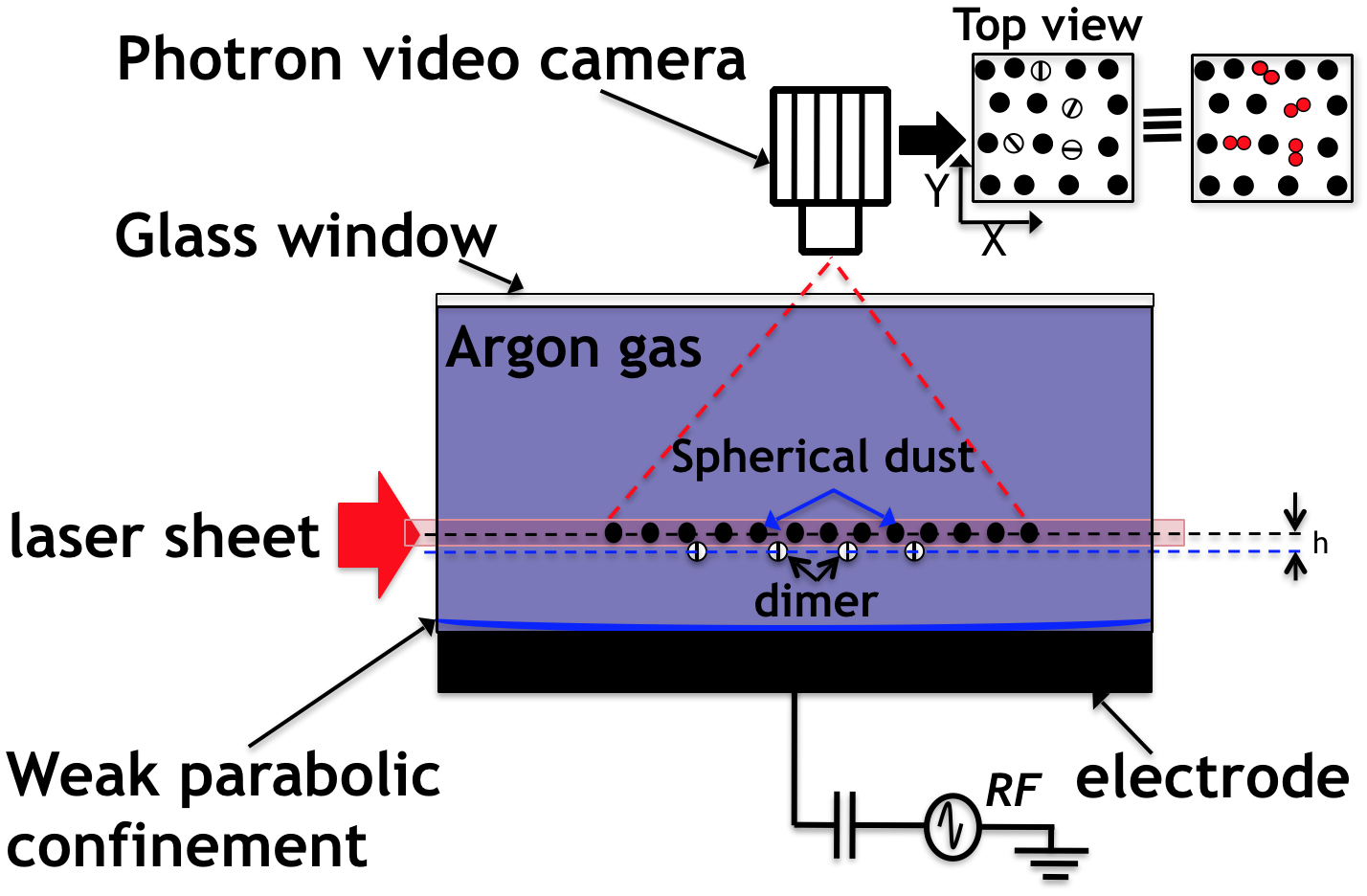}
\caption{Sketch of the experimental conditions to prepare quasi-two-dimensional suspension of spherical dust particles and binary agglomerates in the background of weakly ionized plasmas. The microparticles are trapped in the weak parabolic confinement potential above the rf electrode and are illuminated with a horizontal laser sheet. Unlike spherical particles, the binary agglomerates are identified with rotational interference fringe pattern on their defocus images (marked as striped particle on top view image). The spherical particles and binary agglomerates levitate at different heights with separation `h' as shown.}
\label{GEC} 
\end{figure} 

\begin{figure}
\includegraphics[width=\linewidth]{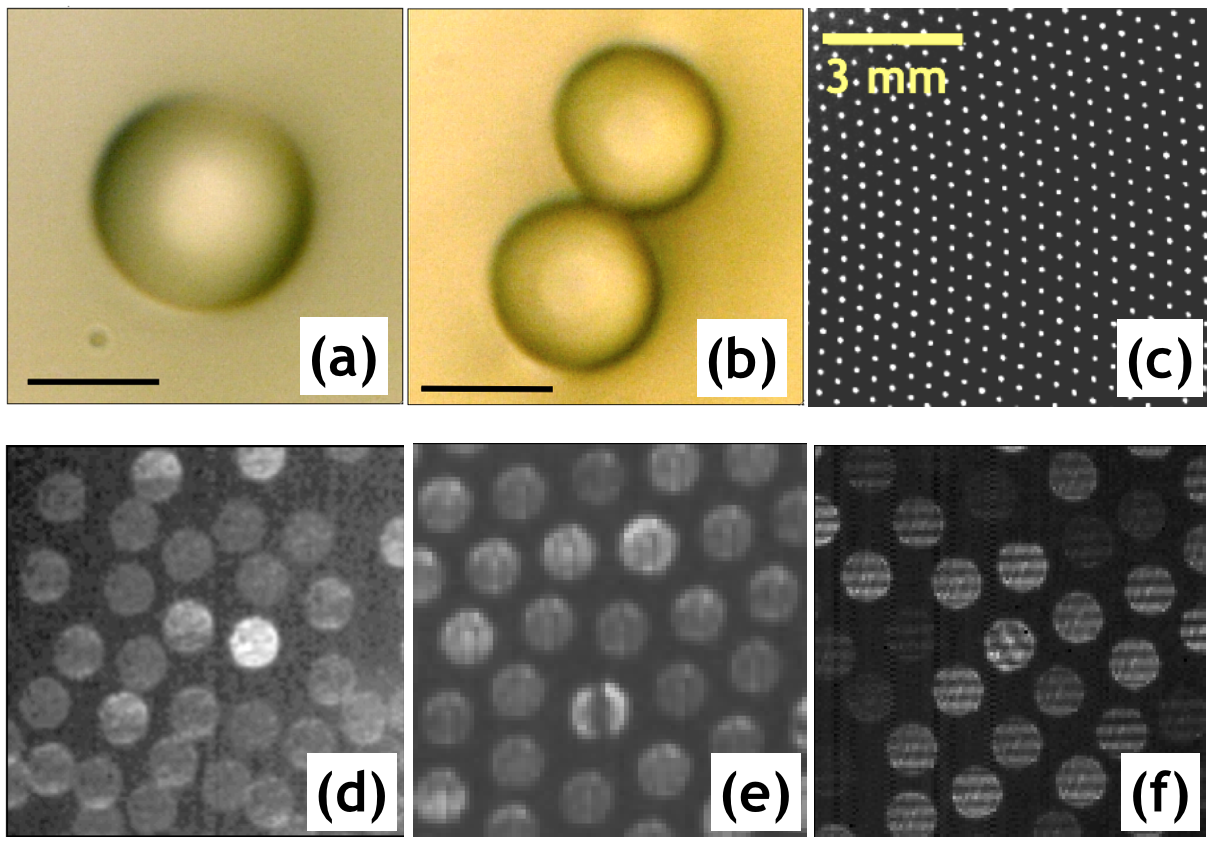}
\caption{(a) Spherical dust particle and (b) binary agglomerate as observed through optical microscope. Scale bar is 5 $\mu$m. (c) Two dimensional plasma crystal made of spherical particles as observed in a focused image using video microscopy. (d) Defocused image of small particle ($\sim$ 4.32 $\mu$m) where no fringe patterns are observed neither on spherical particles nor on binary agglomerates. (e) Defocused images of medium size particles ($\sim$ 7.16 $\mu$m) where rotational fringe patterns are observed on binary agglomerates but not on spherical particles. (f) Large particles ($\sim$ 20 $\mu$m)where stationary fringe patterns are observed on spherical particles and a combination of rotational fringe on top of stationary fringe patterns are observed on their binary agglomerates.}
\label{stripe-particles} 
\end{figure}

Recently defocus imaging technique has been used as an useful diagnostic to identify binary agglomerates in complex plasma which contains rotating interference fringe patterns on their defocus images~\cite{APL_Manis}. Now, it is discovered that stationary interference fringes appear on individual, bigger size, spherical particles. In such cases, a combination of rotational and stationary fringe patterns are the characteristic of binary agglomerates as shown in Fig.~\ref{stripe-particles}f. At some point, the rotational fringe pattern overlaps exactly on top of stationary fringe patterns which implies that inter-fringe spacings are identical on defocus images for such spherical particles and their binary agglomerates. The goal of this work is to put forward an idea of using defocus imaging technique to measure diameter of individual spherical dust particle instantly during experiments within some accuracy. We have tried to explore the origin of the fringe patterns for individual spherical dust particle which itself acts as an efficient interferometer. The fringe pattern becomes distinct as the particle diameter increases and there is a lower cut-off ($\sim$ 9 $\mu$m) below which we don't observe any such fringes. For medium size particles, rotational fringes appear on binary agglomerates but very faint fringes appear on defocus images of individual spherical particles. For smaller particles (below $\sim$ 5.5 $\mu$m) no fringes are observed at all on either spherical particles or on their binary agglomerates. 

%{\color{red}   ILIDS is suitable to measure spatial distribution of the size for transparent spherical droplets and the particle size is determined by the distance between the two glare points~\cite{Glover1995,Maeda2000}. The point spread function of defocused optics is circular and the radii of the glare points from a particle's reflection and refraction become larger. They overlap with each other on the defocused plane which produces parallel fringes in the overlapped area. The characteristic interferogram can be observed with a far-field arrangement of receiving optics. It is important to note that the number of fringes or fringe spacing is related to the particle diameter. The radius of the spread circular image with fringes is not affected by the diameter of particle. The polarization properties of light has also been used to determine the droplet size~\cite{Massoli1989}. Different other techniques have been used to measure the diameters of the particles~\cite{}. For example, the shadow Doppler velocimetry technique has been used for non-spherical and non-transparent particles~\cite{}. Similarly, the phase doppler anemometry~\cite{} was used to measure the diameter of particles by noting the phase differences between the Doppler signals from multiple receivers. }
 
%{\color {blue} (1) camera is at 90 degree (2) 2D dusty plasma with individual particle illumination}

The experiments were performed with a (modified) Gaseous Electronics Conference (GEC) chamber, in a capacitively coupled rf glow discharge at 13.56 MHz (see Fig.~\ref{GEC}). The Argon pressure and the forward rf power were kept at 1 Pa and at 20 Watt respectively. Particles of different sizes and materials have been used for the experiments: Melamine formaldehyde (MF) particles (mass density: 1.51gm/cm$^3$, refractive index (RI): 1.68) with diameters ($2r$) of 4.32 $\mu$m, 7.16 $\mu$m, 8.42 $\mu$m, 9.19 $\mu$m and 14.91$\pm$0.26 $\mu$m; polystyrene (PS) particles (mass density: $\sim$ 1.05 gm/cm$^3$, RI: 1.58) with a diameter of 11.35 $\mu$m and PMMA particles (mass density: $\sim$ 1.19 gm/cm$^3$, RI: 1.49) with diameters of 17.02$\pm$0.03 $\mu$m and 20 $\mu$m. The particle suspension was illuminated with a horizontal sheet of red diode laser light (wavelength of 660 nm) and imaged through the top glass window with the Photron FASTCAM 1024 PCI camera operating at a speed of 60 frames/sec with a field of view of 1024 x 1024 pixels. The focal length of the lens is 105 mm with aperture range, $f/2.8$ to $f/32$. The camera lens was equipped with a narrow-band interference filter to collect only the illumination laser light scattered by the particles.

%\begin{figure}
%\includegraphics[width=\linewidth]{fig5}
%\caption{(a) Schematic diagram of the observed phenomena. The dust particle is illuminated by the laser light and its focussed and defocus images have been visualized by a video camera placed perpendicular to the laser beam. The fringe pattern appears on defocus images due to the interference of reflected and first order refracted light at $90^o$ scattering angle. (b) Schematics of ray diagram of reflected and first order refracted lights within the dust particle. (c) The variation of the incident (red circles) and refracted angles (blue squares) w.r.t refractive indices of the material of the dust particles~\cite{Pajot1999}. The positions of the three perpendicular dash lines represent the refractive indices of three different types of dust particles used in the experiment: MF (RI: 1.68); PS (RI: 1.58) and PMMA (RI: 1.49). It is shown that for all types of particles, the angle of incidence is very small.}
%\label{schematic} 
%\end{figure}
\begin{figure}
\includegraphics[width=0.9\linewidth]{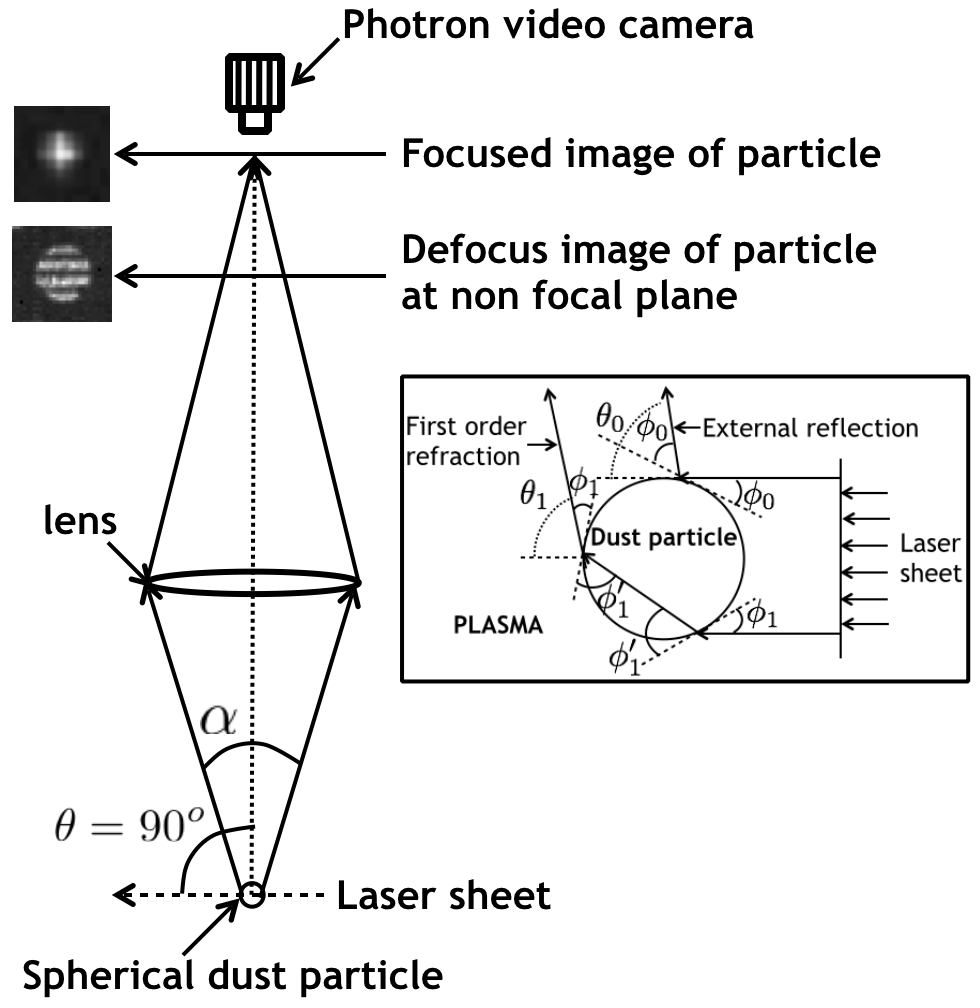}
\caption{(a) Schematic diagram of the observed phenomena. The dust particle is illuminated by the laser light and its focussed and defocused images have been visualized by a video camera placed perpendicular to the laser beam. The fringe pattern appears on defocused images due to the interference of reflected and first order refracted light at $90^o$ scattering angle. Schematics of ray diagram of reflected and first order refracted lights within the dust particle are shown in the inset.}
\label{schematic} 
\end{figure}

When injected in the plasma, both the spherical dust particles and their binary agglomerates become highly charged and form a quasi-two dimensional suspension above the lower electrode~\cite{APL_Manis}. The binary agglomerates levitate just below the monolayer of spherical particles without forming vertical pairs so that all the particles can be viewed from top view camera as shown schematically in Fig.~\ref{GEC}. All the particles can be identified by few bright pixels in a focused image due to laser light scattering. It is not possible to characterize the particle shape and size by looking at these focused images. But as we defocus the images, interesting new features are observed: distinct interference fringe patterns appear on the defocused images of the particles~\cite{APL_Manis}. Identical fringe patterns are observed for particles with same diameter as shown in Fig.~\ref{stripe-particles}f. As we increase the particle size, the number of fringes also increases on the defocused image of a single spherical particle and they become distinct. The observed phenomena {\it i.e.} the appearance of stationary fringe pattern on bigger size, spherical dust particles in plasma environment has been explained in the framework of ``Interferometric Laser Imaging (ILI)'' technique which is based on ``Mie scattering theory'' and takes into account the interference of the scattered light from a single transparent particle. The reflected and first order refracted rays interfere with each other to generate fringe patterns at the defocus plane. This technique has been applied before for measuring size of drops and bubbles (Interferometric Laser Imaging for Droplet Sizing (ILIDS)) in spray dryer systems, spark ignition engine, etc. as mentioned in Ref.~\cite{Maeda2000} and references there in. Two glare points due to reflection and refraction from diametrically opposite positions can be observed at the focal plane if $d >$ 50 $\mu$m. However, for $d <$ 50 $\mu$m, ILI is the most suitable technique to determine particle size. It is to be noted that there is a lower limit of particle diameter below which ILI is invalid: $d_{min} \sim 20\lambda/\pi$ where $\lambda$ is the wavelength of illumination laser. In our experiment, we use $\lambda \sim 660$nm and hence $d_{min} \sim$ 4.20 $\mu$m. 
\begin{figure}
\includegraphics[width=0.9\linewidth]{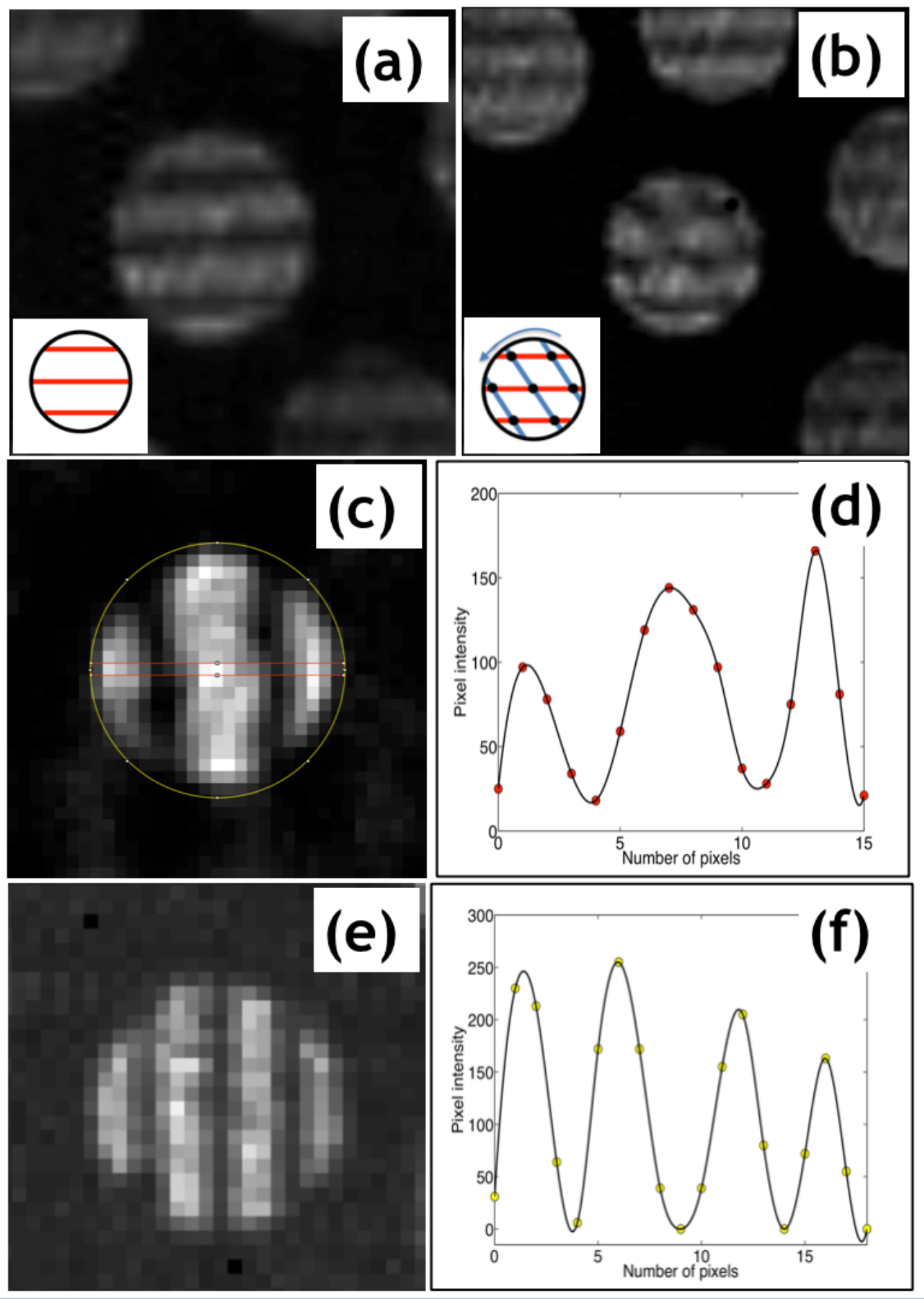}
\caption{(a) Overlap of rotational and stationary fringe patterns for a binary agglomerates. It shows that inter-fringe spacings are same for both types of rings patterns. (b) Non-overlap fringe orientation for binary agglomerates where rotational fringes are oblique w.r.t stationary horizontal fringe patterns. The intersection of these two types of fringes form local dark patches on defocused images as illustrated in the inset by the black dots. (c) Illustration to calculate number of fringes for 11.35 $\mu$m particle. In this case we consider rotational fringe patterns with region of interest across the diameter and perpendicular the fringe orientation. Intensity variation along the ROI has been observed where black pixels correspond fringe position. Fringe separation is determined as the distance between two maxima (or minima) of the intensity distribution. The number of fringes has been obtained by dividing the diameter of the particle (length of ROI) with fringe separation. (d) Intensity variation for 11.35 $\mu$m diameter particle. Similar calculations have been performed for the bigger 20 $\mu$m diameter particles where stationary fringes are formed as shown in (e) and (f). The fringe positions have been flipped to make clear visual effects. It is to be noted that the rotational interference fringe shape for binary agglomerate is not ``exactly'' vertical as observed in stationary fringes for spherical particles which can be due to morphological effect. }
\label{image-patterns} 
\end{figure}
To calculate the number of fringes observed on the defocused image of a single particle, we select a one pixel width horizontal region of interest (ROI) along the diameter at the centre of the particle image. The dark fringes are perpendicular to the ROI. The intensity variation along the ROI exhibits several maxima and minima. As the particle size increases, the number of fringes increases and hence number of maxima/minima increases: N $\sim$ 1.16 for 7.16 $\mu$m, N $\sim$ 1.58 for 9.19 $\mu$m, N $\sim$ 2.07 for 11.35 $\mu$m, N $\sim$ 2.28 for 14.91 $\mu$m, N $\sim$ 2.88 for 17.02 $\mu$m, and N $\sim$ 3.43 for 20 $\mu$m. It is to be noted that with increasing particle size, the width of each peak decreases and height increases indicating distinct as well as sharp features of fringe patterns. It is difficult to measure the fringe separation for smaller particles due to the wider width of fringes and they appear almost at two ends of ROI.

\begin{figure}
\includegraphics[width=0.95\linewidth]{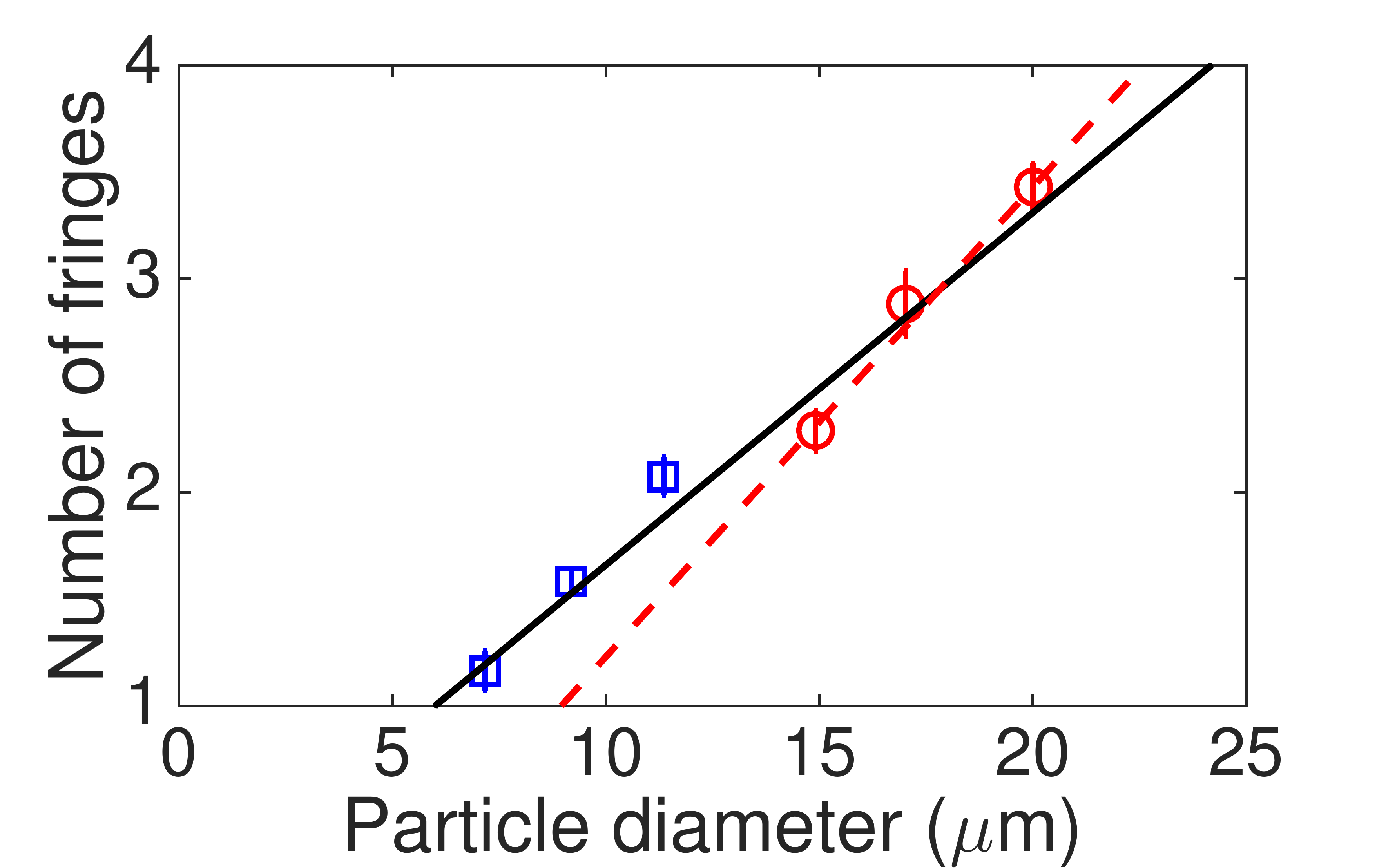}
%\caption{Number of fringes (N) on defocus images of particles increases with particle diameter ($d$). Red circles represent the measurements using stationary fringe patterns on spherical particles. The red dash line represents linear fit with the data for stationary fringe patterns, N = $0.22d - 0.97$, which provides a lower limit $d_c \sim 9 \mu$m below which no stationary fringe pattern on spherical particle is observable. Blue squares represent the measurement using fringe pattern on binary agglomerates. The blue dash line represents linear fit with the data for rotational fringe patterns, N = $0.22d - 0.4$, which provides a lower limit $d_l \sim 6.3 \mu$m below which no rotational fringe patterns are observable. The solid line represents linear fit using all data points combining defocus image analysis of spherical particles and binary agglomerates, N = $0.17d + 0.01$, which provides an estimate of critical diameter of particle $d_c \sim 5.8 \mu$m below which no fringe pattern on binary agglomerates is observable. The discrepancy is $\sim 8\%$ from the binary agglomerates.}
\caption{Number of fringes (N) on defocused images of particles increases with particle diameter ($d$). Red circles represent the measurements using stationary fringe patterns on spherical particles. The red dash line represents linear fit with the data and provides a lower limit $d_c \sim 9 \mu$m below which no stationary fringe pattern on spherical particle is observable. Blue squares represent the measurement using fringe pattern on binary agglomerates. The solid line represents the best linear fit using all data points combining defocused image analysis of spherical particles and binary agglomerates, N = $0.17d + 0.01$, which provides an estimate of critical diameter of particle $d_c \sim 5.8 \mu$m below which no rotational fringe pattern on binary agglomerates is observable. }
\label{dimer-rotation}
\end{figure}

According to Lorentz-Mie theory, the light scattered by a spherical particle is inhomogeneously distributed in space (oscillating function of the angle in the range $0<\theta<\pi$) which depends on particle diameter, refractive index and incident light characteristics~\cite{hulst}. The origin of these oscillations is due to interference between reflected, refracted and diffracted rays coming out of the particle and forms the basis of the Mie Scattering Interferrometry. However, for bigger particle it was shown that simpler geometric analysis can be used as an alternate of complex Mie theory to estimate particle size for a scattering angle centred around $90^o$. 
To analyze the phenomenon, we consider all dust particles are perfectly spherical and homogeneous. The interaction between laser beam and the particle is shown in Fig.~\ref{schematic}. The total scattering light intensity is due to the sum of reflection and first order refraction rays. 
%The reflection is not attenuated by absorption but the first order refracted ray is: the optical path difference is a function of particle diameter and is responsible for dark and bright fringes. 
%The angles of incidence for the reflected and refracted beams are $\phi_0$ and $\phi_1$ respectively . If $\phi_1'$ is the angle of refraction then Snell's law is: $cos\phi_1' = \frac{1}{m}cos\phi_1$. For pure reflection and for first order refraction, the scattering angles from the incident ray are $\theta_0 = 2\phi_0$ and $\theta_1 = 2(\phi_1' - \phi_1)$ respectively. There exists only two rays for which these scattering angles are identical. It can be shown that the value of $\phi_1$ is very low for a scattering angle of $90^o$ and typical values of refractive indices of dust particles. Only very few rays can be reflected once and leave the particle with a $90^o$ scattering angle. 
The phase difference between the reflected and refracted rays can be expressed as~\cite{Pajot1997,Golombok1998}:
\bee
\phi_0 - \phi_1 = \frac{2\pi d}{\lambda}\lt(sin\frac{\theta}{2} - \sqrt{m^2 + 1 - 2m cos\frac{\theta}{2}}\rt)
\ene
An infinitesimal variation of the scattering angle induces a maximum or minimum light intensity variation when the phase difference is equal to $2\pi$. %Hence we can write,
%%\bee
%%\Delta(\delta_0 - \delta_1) = \frac{2\pi d}{\lambda}\lt(cos\frac{\theta}{2} + \frac{m sin\frac{\theta}{2}}{\sqrt{m^2 + 1 - 2m cos\frac{\theta}{2}}}\rt)\Delta\theta = 2\pi
%%\ene
So, the angular inter-fringe spacing $\Delta\theta$ can be related to the particle diameter:
\bee
\Delta\theta = \frac{2\lambda}{d}\lt(cos\frac{\theta}{2} + \frac{m sin\frac{\theta}{2}}{\sqrt{m^2 + 1 - 2m cos\frac{\theta}{2}}}\rt)^{-1}
\ene
If the scattering angle is of $90^o$ then it can be assumed that the incidence angle of the refracted ray on the particle is close to zero and hence the above equation can be simplified as,
\bee
\Delta\theta = \lt(\frac{2\lambda}{d}\rt)\frac{1}{1+\frac{1}{m}}
\label{deltatheta}
\ene
The number of fringes on the defocused image of a spherical particle depends on the collection angle, $\alpha$ which is equal to the product of number of fringes $N$ and angular fringe spacing, $\Delta\theta$:
\bee
d = \lt(\frac{2\lambda N}{\alpha}\rt)\frac{1}{1+\frac{1}{m}}
\ene
%The collection angle can further be approximated as function of the magnification ratio, G (ratio of the image size and the object size) and the numerical aperture, N.A:
%\bee
%\alpha \sim 2sin^{-1}\lt(\frac{G}{G+1}.\frac{1}{2N.A}\rt)
%\label{alpha}
%\ene
%Combing Eqns.\ref{delta theta} and \ref{delta}, the particle diameter can be expressed as:
%\bee
%d = \frac{\lambda N(1+\frac{1}{m})^{-1}}{sin^{-1}\lt(\frac{G}{G+1}.\frac{1}{2N.A}\rt)}
%\ene

The results have been described in Table-1 in which the left most column represents the diameter of the particles as specified by the manufacturer. Then Eqn. 4 has been used to estimate the particle diameters in experiments for known refractive indices of materials, wavelength of the illuminating lasers and specified diameters (14.91, 17.02 and 20 $\mu$m).  

It is to be noted that the diameters for smaller particles (7.16, 9.19 and 11.35 $\mu$m) have been estimated by analyzing rotational interference fringes on defocused images of their binary agglomerates. It is based on the conjecture that the same measurement technique as described above for spherical particles is also applicable for defocused image analysis of binary agglomerates which contain rotational fringe patterns. This is due to the fact that the inter-fringe spacing for rotational pattern and stationary patterns are same for bigger size particles where both patterns are visible. This simplified approximation agrees well with experimental observations. However, the full understanding of the source of these rotating fringes is still unknown and kept for future work, but they certainly represent the dynamic signatures of binary agglomerates. It is found that the estimated diameters are sufficiently close to those of specified diameters with maximum tolerance of $\sim$ 14\% for 11.35 $\mu$m particles and  minimum of $\sim$ 2\% for 14.91 $\mu$m particles.

 \begin{table}[h]
\begin{center}
    \begin{tabular}{|c|c|c|c|c|c|}
    \hline
    d ($\mu$m) & Material & RI & Defocused & N: \# of  & d ($\mu$m) \\
    Specified & & (m) & image &fringes & Estimated  \\
     & & & analysis & & \\ \hline
    7.16 & MF & 1.68 & BA &1.16 $\pm$ 0.09 & 7.43 $\pm$ 0.64  \\ \hline
    9.19 & MF & 1.68 & BA &1.58 $\pm$ 0.05 & 10.12  $\pm$ 0.32 \\ \hline
    11.35 & PS & 1.58 & BA &2.07 $\pm$ 0.09 & 12.90 $\pm$ 0.50 \\ \hline
    14.91 & MF & 1.68 & SP, BA&2.28 $\pm$ 0.09 & 14.60  $\pm$ 0.52 \\ \hline
    17.02 & PMMA & 1.49 & SP, BA &2.88 $\pm$ 0.15 & 17.57  $\pm$ 0.89 \\ \hline
    20 & PMMA & 1.49 & SP, BA& 3.43 $\pm$ 0.10 & 20.91  $\pm$ 0.61 \\ \hline  
    \end{tabular} 
\caption{Particle diameter has been estimated using Eqn.4 by counting the number of interference fringes (N) on defocused images of different spherical particles (SP) or binary agglomerates (BA) of different refractive indices (m) but with same laser wavelength ($\lambda \sim 660$nm) and collection angle ($\alpha \sim 23^o$)}
    \label{table1}
\end{center}
\end{table}

In conclusion, we have discussed a simple and useful method to estimate size of a spherical particle over a wide size range by analyzing defocused images of both spherical particles and their binary agglomerates. The diameter of the spherical particle has been estimated for the first time by counting the number of interference fringes and their separation in the framework of interferometric laser imaging methods. The stationary fringe pattern is distinct for bigger spherical particles but they are not clearly visible for medium size particles. To overcome this problem, the separation of the rotating fringes for the binary agglomerates has been used in this size range. The number of fringes increases with particle size and there exists two critical diameters below which we do not observe any stationary and rotational fringe patterns. This simple technique can be used to identify size and shape of impurities or polydispersities in laboratory experiments as well as under microgravity conditions.\\

* Email: manischaudhuri@g.harvard.edu \\
{\bf Acknowledgement:} M.C is supported by Marie-Curie international outgoing fellowship.


\begin{thebibliography}{99}

\bibitem{Chaikin:Lubensky} P. M. Chaikin and T. Lubensky, {\it Principles of Condensed Matter Physics} (Cambridge University Press, Cambridge, 2000).
\bibitem{kittel} C. Kittel, {\it Introduction to Solid State Physics} (Wiley, 2004).
\bibitem{Hansen:McDonald} J.-P. Hansen and I. McDonald, {\it Theory of Simple Liquids} (Academic Press, 2013).
\bibitem{larson} R. G. Larson, {\it The Structure and Rheology of Complex Fluids} (Oxford University Press, 1998).
\bibitem{lowen1994} H. L\"owen, Phys. Rep. {\bf 237}, 249324 (1994).
\bibitem{biben1994} T. Biben, R. Ohnesorge, and H. L\"owen, Europhys. Lett. {\bf 28}, 665670 (1994).
\bibitem{Weeks:2000aa} E. R. Weeks, J. C. Crocker, A. C. Levitt, A. Schofield, and D. A. Weitz, Science {\bf 287}, 627 (2000).
\bibitem{Gasser:2001aa} U. Gasser, E. R. Weeks, A. Schofield, P. N. Pusey, and D. A. Weitz, Science {\bf 292}, 258 (2001).
\bibitem{Dullens2006} R. P. A. Dullens, D. G. A. L. Aarts, and W. K. Kegel, Phys. Rev. Lett. {\bf 97}, 228301 (2006).
\bibitem{Kawasaki:2010aa} T. Kawasaki and H. Tanaka, PNAS {\bf 107}, 14036 (2010).
\bibitem{Peng-Tan:2014aa} N. X. Peng Tan and L. Xu, Nat. Phys. {\bf 10}, 73 (2014).
\bibitem{Morfill_RMP} G. E. Morfill and A. V. Ivlev, Rev. Mod. Phys. {\bf 81}, 1353 (2009).
\bibitem{Chaudhuri_SM} M. Chaudhuri, A. V. Ivlev, S. A. Khrapak, H. M. Thomas, and G. E. Morfill, Soft Matter. {\bf 7}, 1229 (2011).
\bibitem{Morfill:lowen} A. Ivlev, H. L\"owen, G. Morfill, and C. P. Royall, {\it Complex plasmas and colloidal dispersions: Particle-resolved studies of classical liquids and solids} (World Scientific, Singapore, 2012).
\bibitem{Thomas_PRL} H. M. Thomas, G. E. Morfill, and et. al, Phys. Rev. Lett. {\bf 73}, 652 (1994).
\bibitem{Chu_PRL} J. H. Chu and L. I, Phys. Rev. Lett. {\bf 72}, 4009 (1994).
\bibitem{Thomas-Nature-1996} H. M. Thomas and G. E. Morfill, Nature {\bf 379}, 806 (1996).
\bibitem{Fortov2005PR}V. E. Fortov, A. V. Ivlev, S. A. Khrapak, A. G. Khrapak, and G. E. Morfill, Phys. Rep. {\bf 421}, 1 (2005).
\bibitem{vladimirov-ost-phys.rep-2004} S. V. Vladimirov and K. Ostrikov, Phys. Rep. {\bf 393}, 175 (2004).
\bibitem{Konopka2000} U. Konopka, G. E. Morfill, and L. Ratke, Phys. Rev. Lett. {\bf 84}, 891 (2000).
\bibitem{lampe-gos-ste-phys.plasmas-2003} M. Lampe, R. Goswami, Z. Sternovsky, S. Robertson, V. Gavrishchaka, G. Ganguli, and G. Joyce, Phys. Plasmas {\bf 10}, 1500 (2003).
\bibitem{Chaudhuri2007a} M. Chaudhuri, S. A. Khrapak, and G. E. Morfill, Phys. Plasmas {\bf 14}, 022102 (2007).
\bibitem{Chaudhuri2008a} M. Chaudhuri, S. A. Khrapak, and G. E. Morfill, Phys. Plasmas {\bf 15}, 053703 (2008).
\bibitem{ChaudhuriIEEE2010} M. Chaudhuri, S. A. Khrapak, R. Kompaneets, and G. E. Morfill, IEEE Trans. on Plasma Sci. {\bf 38}, 818 (2010).
\bibitem{Morfill:pop} G. E. Morfill, A. V. Ivlev, and H. M. Thomas, Phys. Plasmas {\bf 19}, 055402 (2012).
\bibitem{Quinn_PhysRevE.64.051404} R. A. Quinn and J. Goree, Phys. Rev. E {\bf 64}, 051404 (2001).
\bibitem{Zuzic:PhysRevLett.85.4064} M. Zuzic, A. V. Ivlev, J. Goree, G. E. Morfill, H. M. Thomas, H. Rothermel, U. Konopka, R. S\"utterlin, and D. D. Goldbeck, Phys. Rev. Lett. {\bf 85}, 4064 (2000).
\bibitem{Nosenko_PhysRevLett.93.155004} V. Nosenko and J. Goree, Phys. Rev. Lett. 93, 155004 (2004).
\bibitem{Nosenko_PhysRevLett.103.015001} V. Nosenko, S. K. Zhdanov, A. V. Ivlev, C. A. Knapek, and G. E. Morfill, Phys. Rev. Lett. {\bf 103}, 015001 (2009).
\bibitem{Hartmann_PhysRevLett.105.115004} P. Hartmann, A. Douglass, J. C. Reyes, L. S. Matthews, T. W. Hyde, A. Kovacs, and Z. Donko, Phys. Rev. Lett. {\bf 105}, 115004 (2010).
\bibitem{Feng_PhysRevLett.104.165003} Y. Feng, J. Goree, and B. Liu, Phys. Rev. Lett. {\bf 104}, 165003 (2010).
\bibitem{Nosenko_PhysRevLett.99.025002} V. Nosenko, S. Zhdanov, and G. Morfill, Phys. Rev. Lett. {\bf 99}, 025002 (2007).
\bibitem{Knapek_PhysRevLett.98.015004} C. A. Knapek, D. Samsonov, S. Zhdanov, U. Konopka, and G. E. Morfill, Phys. Rev. Lett. {\bf 98}, 015004 (2007).
\bibitem{Nunomura_PhysRevLett.95.025003} S. Nunomura, D. Samsonov, S. Zhdanov, and G. Morfill, Phys. Rev. Lett. {\bf 95}, 025003 (2005).
\bibitem{Liu_PhysRevLett.100.055003} B. Liu and J. Goree, Phys. Rev. Lett. {\bf 100}, 055003 (2008).
\bibitem{Juan_PhysRevE.64.016402} W.-T. Juan, M.-H. Chen, and L. I, Phys. Rev. E {\bf 64}, 016402 (2001).
\bibitem{Morfill:PhysRevLett.83.1598} G. E. Morfill, H. M. Thomas, U. Konopka, H. Rothermel, M. Zuzic, A. Ivlev, and J. Goree, Phys. Rev. Lett. {\bf 83}, 1598 (1999).
\bibitem{Kretschmer:PhysRevE.71.056401} M. Kretschmer, S. A. Khrapak, S. K. Zhdanov, H. M. Thomas, G. E. Morfill, V. E. Fortov, A. M. Lipaev, V. I. Molotkov, A. I. Ivanov, and M. V. Turin, Phys. Rev. E {\bf 71}, 056401 (2005).
\bibitem{Zuzic:Naturephysics} M. Rubin-Zuzic, G. E. Morfill, A. V. Ivlev, R. Pompl, B. A. Klumov, W. Bunk, H. M. Thomas, H. Rothermel, O. Havnes, and A. Fouquet, Nat. Phys. {\bf 2}, 181185 (2006).
\bibitem{Thomas:NJP} H. M. Thomas, G. E. Morfill, V. E. Fortov, A. V. Ivlev, V. I. Molotkov, A. M. Lipaev, T. Hagl, H. Rothermel, S. A. Khrapak, R. K. Suetterlin, M. Rubin-Zuzic, O. F. Petrov, V. I. Tokarev, and S. K. Krikalev, New Journal of Physics {\bf 10}, 033036 (2008).
\bibitem{Schwabe:EPL} M. Schwabe, K. Jiang, S. Zhdanov, T. Hagl, P. Hu- ber, A. V. Ivlev, A. M. Lipaev, V. I. Molotkov, V. N. Naumkin, K. R. Stterlin, H. M. Thomas, V. E. Fortov, G. E. Morfill, A. Skvortsov, and S. Volkov, Europhysics Letters {\bf 96}, 55001 (2011).
\bibitem{Ed_Thomas_10.1063/1.873544} E. Thomas, Phys. Plasmas {\bf 6}, 2672 (1999).
\bibitem{Ed_Thomas_10.1063/1.2174831} E. Thomas and J. Williams, Phys. Plasmas {\bf 13}, 055702 (2006).
\bibitem{Kroll_10.1063/1.2932109} M. Kroll, S. Harms, D. Block, and A. Piel, Phys. Plasmas {\bf 15}, 063703 (2008).
\bibitem{Annaratone_PPCF} B. M. Annaratone, T. Antonova, D. D. Goldbeck, H. M. Thomas, and G. E. Morfill, Plasma Physics and Controlled Fusion 46, B495 (2004).
\bibitem{kading:090701} S. Kading and A. Melzer, Physics of Plasmas {\bf 13}, 090701 (2006).
\bibitem{APL_Manis} M. Chaudhuri, V. Nosenko, C. Knapek, U. Konopka, A. V. Ivlev, H. M. Thomas, and G. E. Morfill, Applied Physics Letters 100, 264101 (2012).
\bibitem{Maeda2000} M. Maeda, T. Kawaguchi, and K. Hishida, Meas. Sci. Technol. {\bf 11}, L13 (2000).
\bibitem{hulst} H. V. de Hulst, {\it Light scattering by small particles} (Wiley, 1957).
\bibitem{Pajot1997} C. Mounmim-Rousselle and O. Pajot, Optical Technology in Fluid, Thermal and Combustion III {\bf 3172}, 700 (1997).
\bibitem{Golombok1998} M. Golombok,V. Morin, and C. Mounaim-Rousselle, J. Phys. D: Applied Physics {\bf 31}, pL59 (1998)

\end{thebibliography}
\end{document}